**Multiferroic double perovskites: Opportunities, Issues and Challenges**

**M.P. Singh, K.D. Truong, S. Jandl, and P. Fournier**
**Regroupement québécois sur les matériaux de pointe, Département de physique,**
**Université de Sherbrooke, Sherbrooke, J1K 2R1 Canada**

**Abstract**

Despite the great technological and scientific importance, a well-controlled coupled behavior in magnetoelectric multiferroic materials at room temperature remains challenging. We demonstrate that the self-ordered $A_2B'B''O_6$ (e.g., A = La, B' = Mn, and B'' = Ni) double perovskites provide a unique alternative opportunity to control and/or to induce such behavior in oxides. In this paper, we outline and discuss the various challenges and bottleneck issues related to this class of materials using $La_2NiMnO_6$ as a prototype. Details of the synthesis strategies, using pulsed laser deposition as a central tool, to obtain a meticulous control over B'/B'' cation ordering is illustrated. The possible impact of Ni/Mn cation ordering on the magnetic ground states, ferromagnetic transition temperature, phonon behavior, and spin-phonon coupling are presented. Our results are also compared with the current literature on related compounds.

# I. Introduction

Magnetoelectric multiferroics have recently received considerable interest for both technological and fundamental motivations[1]. The simultaneous presence of ferromagnetism and ferroelectricity in a single phase, and the possible spin-phonon and spin-polar couplings in these systems offer the appealing opportunity to design various unconventional devices, such as multiple-state memory elements and electric-field controlled magnetic sensors. However, their ultimate uses depend upon the nature and net amplitude of the coupling around 300 K. The coexistence of various electronic order parameters that can simultaneously interact in a given system also poses new challenges for physics, chemistry, and technology [1]. In this context, there are various stumbling blocks. The most challenging ones are how to design the materials with well controlled spin-polar and spin-phonon couplings? How to promote and to control these couplings around room temperature? Answers to these questions are hence crucial for designing materials with well-controlled coupled behaviour in real devices and for the understanding of the microscopic mechanism behind the multi-ferroic character.

Recently, various avenues have been used and/or proposed to address these issues. They include the use of spin frustration as a potential source for inducing the electrical polarization in a given system owing to the strong-spin orbit coupling and Dzyaloshinski-Moria (DM) interactions [2]; the design of superlattices [3] and/or composites [4] based on magnetic and ferroelectric materials. These approaches are indeed successful, but often lead to either a weak ferromagnetic behavior and/or to the presence of coupling effects below liquid nitrogen temperature that limits their applicability. In this context, double perovskites $A_2B'B''O_6$ (where B' has partially filled $e_g$-orbitals whereas B'' has empty $e_g$-orbitals or *vice-versa*) provide a unique alternate opportunity to promote multiferroic behaviours because they are ferromagnetic insulators [5] owing to the 180°-superexchange interactions between B'/B'' cations *via* the oxygen ions [6-9]. In this respect, the demonstration of a magnetodielectric effect in the vicinity of the ferromagnetic-to-paramagnetic transition in bulk $La_2NiMnO_6$ [6] and $La_2CoMnO_6$ films [7] are indeed a step forward. However, the magnetodielectric effect in $La_2NiMnO_6$ thin films [8] is observed far below their magnetic transition temperature. One of the possible reasons for this different behaviour may be related to the short-range Ni/Mn order in these films, as demonstrated by Raman scattering. In LNMO thin films [10], it was recently shown that such structural order is limited only to small domain sizes of about 50-100 nm.

The above mentioned studies suggest that the B'/B'' ordering in $A_2B'B''O_6$ is crucial in determining their functional properties. The most important challenges are therefore how to control and to achieve the B'/B'' structural ordering in double perovskites and how their functional properties are correlated. To address these aforementioned issues, we have studied $La_2NiMnO_6$ epitaxial films grown by pulsed laser deposition. In this paper, we present the phase stability, structure, magnetic and spin-phonon coupling behaviors of LNMO films with varying degrees of Ni/Mn ordering. We also compare our results with our previous work and the published literature on related double perovskites, *viz.,* $La_2CoMnO_6$ (LCMO) and $Pr_2NiMnO_6$ (PNMO) films.

## II. Experimental details

For ablation, a polycrystalline stoichiometric LNMO target was synthesized using the standard solid-state chemistry route. Epitaxial LNMO films were grown on (111)-$SrTiO_3$ (STO) substrates in a temperature range from 475 to 850 $^{o}$C and under an oxygen pressure ranging from 100 to 900 mTorr using the pulsed-laser deposition (PLD) technique. Following the ablation, the growth chamber was filled with 400 Torr of $O_2$ and the films were cooled down to room temperature at a rate of 10$^{o}$C/min. The crystallinity and the coherent nature of our films were checked using x-ray diffraction. Raman backscattering measurements with a 0.5 $cm^{-1}$ resolution using a CCD camera were carried out between 10 and 295 K in the XX, XY, and X'Y' and X'X' scattering configurations. The temperature and field dependence of the magnetization were measured using a Superconducting Quantum interferometer device (SQUID) magnetometer from Quantum Design.

## III. B'/B'' cation ordering: *a general perspective*

An ideal perovskite with a chemical formula $AB'O_3$ (*e.g.,* A = La, and B'= Mn) can be described by a primitive cubic unit cell with the A-cation located at its centre and the B-cation on the corners and surrounded by oxygen atoms in an octahedral coordination [11]. In real materials, a deviation from the primitive cubic unit cell is expected due to the differences in the A, B', and oxygen ion sizes, together with their electronic, oxidation and magnetic states. This may ultimately induce lower crystal symmetry and is easily interpreted using the tolerance factor [11]. In particular, $LaMnO_3$ exhibits orthorhombic symmetry whereas $LaNiO_3$ is rhombohedral. Doping of 50% B'' cation at B' sites in $AB'O_3$ may lead to three possible situations. *First*, both B'/B'' cations (Top left panel of Fig. 1) are randomly distributed. In such scenario, the compound is described by a chemical formula of $AB'_{0.5}B''_{0.5}O_3$ and termed as a disordered system. The *second* possible scenario is that B' and B'' cations are alternatively arranged in the unit cell leading to an $A_2B'B''O_6$ formula unit (top-right panel of Fig 1), and termed as ordered double perovskite. Such a cation ordering ultimately leads to a redefinition of the lattice parameters with respect to the pseudo-cubic unit cell. In particular, the ordered LNMO exhibits a monoclinic $P2_1/n$ symmetry [6, 12, 13] with the cell parameters a = 5.515 Å ($a_{P1}$ = a/√2 ~ 3.911Å), b = 7.742 Å ($a_{P2}$ = b/2 ~ 3.871Å), c = 5.46 Å ($a_{P3}$ = c/√2 ~ 3.872Å), and β = 90.04°. The disordered phase, however, possesses an orthorhombic Pbnm symmetry with the lattice parameters a = 5.50 ($a_P$ = a/√2 ~ 3.90Å) Å, b = 7.736 Å ($a_P$ = b/2 ~ 3.868Å), c = 5.54 Å ($a_P$ = c/√2 ~ 3.929 Å)   $a_P$ stands for the lattice parameter of an ideal pseudocubic unit cell.  Most importantly, a disordered LNMO has 4 formula units (f.u.) $LaNi_{0.5}Mn_{0.5}O_3$ *per* unit cell while an ordered LNMO has 2 f.u. $La_2NiMnO_6$ *per* unit cell. The *third* appealing possibility is the intermediate configuration wherein B'/B'' ordering is limited to only a few unit cells or domains as this phase can be defined as short-range ordered. In this particular scenario, there is a possibility of phase separation of both ordered and disordered domains leading to the admixture phase. It is here important to note these three distinct phases are thermodynamical stable. Also of importance, the short-range ordering can also be induced through oxygen or cation vacancy and thus is very sensitive to the synthesis conditions [5, 10]. This outlines the real complexity associated in obtaining a totally ordered phase.

From a material synthesis perspective, the following criteria are useful to obtain the cation ordering in a given system. *First,* a large difference between the ionic size and electronegativity of the B'/B'' cations is crucial to promote the ordering. *Second,* a difference in the oxidation states of the B'/B'' cations will play a crucial role. In the case of LNMO, it is expected that the Ni and Mn ions are respectively in the 2+ and 4+ oxidation states in the ordered phase, while both are in the 3+ oxidation state in the disordered phase. However, further study is warranted to confirm the oxidation states of Ni/Mn in the disordered phase. In fact, the difference in the oxidation states of B' and B'' cations in a self-ordered phase induces a polar character, which will otherwise be absent in the disordered phase. For example, it has recently been shown that the ordered $La_2NiMnO_6$ have the $Ni^{2+}/Mn^{4+}$ cationic configuration leading to large dielectric constants [6]. A similar is also observed for $La_2CoMnO_6$ [18].

## IV. Results and Discussion

The crystallinity and the epitaxial quality of the LNMO films were studied using X-ray diffraction (XRD) in the θ-2θ and rocking (ω) curve modes. Details of the XRD results are published elsewhere [14]. Briefly, as-grown films, irrespective of the growth conditions, always exhibit a diffraction peak in close proximity to the STO (111) reflections. This reveals that the films are grown coherently with 3d-metal interatomic distances matching closely that of STO. Most importantly, the films deposited in a narrow growth window around 800 °C and 800 mTorr display additional XRD reflections. These additional reflections seen only on (111) STO substrates were best interpreted in terms of superlattice reflections resulting from the Ni/Mn ordering in our films [14]. Using these superlattice reflections, we have estimated the corresponding chemical modulation (Λ) length in the ordered LNMO films according to $\Lambda = n\lambda/2[\sin\theta_i - \sin\theta_{i-1}]$, where $\lambda$ is the X-ray wavelength, $\theta_i$ is the peak position of the $i^{th}$ satellite peak, and "n" is the interference order. The average value of Λ was 14.75 Å, a value very close to the theoretical value of 14 Å for an ideal $ABO_3/AB'O_3$ superlattice. This illustrates that a clear B'/B''-site ordering provokes superlattice reflections owing to the transition from high symmetry to a lower one. In LNMO, the orthorhombic symmetry of the disordered phase transforms into the monoclinic structure in the ordered phase.

Using the structural and physical properties of our films as well as those in the literature [7-10, 14-16], we mapped out the phase-stability diagram (Fig. 1) for the Ni/Mn cation ordering in LNMO. Figure 1 clearly shows that strict growth conditions are crucial to obtain either Ni/Mn ordered (circular zone) or Ni/Mn disordered (rectangular zone) LNMO thin films. In particular, the ordered LNMO films were grown in a narrow temperature window of only ±20 $^{o}$C around 810 $^{o}$C and about 800 mTorr. These films have all the expected physical properties of the ordered phase. A deviation from these two growth zones lead to an admixture of both phases (*i.e.,* ordered and disordered) in different proportions. Indeed, the growth window for obtaining the short-range ordered films are broad. The observed phase-stability zone for LNMO is also well-consistent with the phase-stability diagrams of $La_2CoMnO_6$ films [18]. One may further note that the phase-stability zone for the ordered LNMO is narrower than that of LCMO [18]. Our study, thus, illustrates that the ordered B'/B''-site configuration in double perovskite systems can be stabilized by growing the films at relatively high temperatures and under relatively high oxygen pressure. Additionally, the size of the growth zone in the phase diagram may vary for different sets of B'/B'' ions or A-type ions. For example, our recent study on PNMO films show that these films, irrespective of growth conditions used, remain short-range ordered.

Polarized Raman scattering is a stringent measurement to illustrate the B'/B'' cation ordering in double perovskites and has indeed been successfully used to identify local ordering, structural distortions, and spin-phonon coupling [17, 19]. We measured the temperature dependence of the Raman spectra of our disordered, ordered and admixture films in the XX, X'X', XY, X'Y' polarization configurations [19]. Typical Raman spectra of these films, measured at 10 K in the XX and XY configurations, are shown in Figure 2. It shows that the films are characterized by several strong modes. However, the total number of modes and the spectral characteristics (*viz.,* intensity and peak width) **vary** from one sample to the other, confirming the strong influence of Ni/Mn ordering on the observed LNMO phonon excitations. A detailed analysis of these spectra shows that the disordered films are characterized by broad and fewer number of phonon excitations. On contrary, these modes evolve as sharp peaks with comparatively large intensity in the admixture as well as in the ordered films. For example, the anti-stretching 540 cm$^{-1}$ mode in the XX configuration of the disordered films evolves into a well-defined peak in the admixture films (529 cm$^{-1}$) whereas two peaks at 503 cm$^{-1}$ and 527 cm$^{-1}$ respectively are observed for the ordered films. Similar distinctive features in Raman spectra of these films can also be noticed in the XY configuration. As an illustration, the stretching modes at 643 cm$^{-1}$ and 666 cm$^{-1}$ in the disordered films evolve into two clear modes at 648 cm$^{-1}$ and 665 cm$^{-1}$ in the admixture films compared to four well-defined active Raman modes at 621 cm$^{-1}$, 630 cm$^{-1}$, 645 cm$^{-1}$, 665 cm$^{-1}$ in the ordered films. Phonon modes in the disordered films are characterized by relatively large full-widths at half maximum (FWHM ~ 33 cm$^{-1}$) compared to the ordered film (FWHM ~ 23 cm$^{-1}$) at 10 K, confirming the random distribution of the Ni and Mn cations in the disordered films.

The appearance of additional Raman active phonons in the ordered films is due to Brillouin zone folding caused by an increase in the lattice parameters of the ordered phase [19]. Briefly, Raman spectra of a pseudocubic AB'O$_3$ perovskite are characterized by a distinct number of Raman-active phonons and their intensity shows a unique dependence on the polarization configurations. However, doping with B'' cations at the B'-site likely induces local structural distortions that impacts the phonon spectrum in the following ways: If B'/B'' are randomly distributed at the B-sites, as in LaNi$_{0.5}$Mn$_{0.5}$O$_3$, the number of Raman excitations remains the same as that of AB'O$_3$. Nonetheless, it changes the frequency of the phonon excitations and their peak widths because of the changes in the average force constants and the phonon lifetimes. In the ordered A$_2$B'B''O$_6$, the lattice parameters increase which provokes a Brillouin zone folding leading to new Γ-point Raman excitations in the phonon spectra. Thus, the ordering has a major impact on the total number of observed phonons. Experimentally, a clear splitting of Raman-active modes occurs, as seen in the ordered LNMO films, and should be considered as a generic signature of cation ordering in these double perovskites.

Irrespective of the level of Ni/Mn ordering, these films exhibit **a** well-defined magnetic hysteresis loop. The ordered films have about a 4.8 µ$_B$/f.u. saturation magnetization whereas its value reaches about 3.7$\mu_B$/f.u. for the disordered films[14]. The typical temperature dependence of the magnetization (*viz.,* M-T curves) measured in a 500 Oe magnetic field is shown in Figure 3a for the different level of ordering. These data show that both the ordered and disordered films display a single magnetic transition at 270 K and 138 K, respectively. On the contrary, the admixture films display two magnetic transitions (at 295 K and 140 K) illustrating that these films have both ordered and disordered domains. The magnetic interactions in LNMO are governed by the 180°-superexchange process [5] arising from the bond between oxygen and the partially-filled Ni ($e_g$) and Mn ($e_g$) ions. The electronic configurations of Ni and Mn, set by their possible oxidation states $Ni^{2+}$ ($d^8$: $t_{2g}^6 e_g^2$), $Mn^{4+}$ ($d^3$: $t_{2g}^3 e_g^0$), low-spin $Ni^{3+}$ ($d^7$: $t_{2g}^6 e_g^1$) and high spin $Mn^{3+}$ ($d^4$: $t_{2g}^3 e_g^1$) configurations therefore play a crucial role in determining the superexchange strength. In the disordered films, the superexchange interaction for the $Ni^{3+}$-O-$Mn^{3+}$ bonds is likely responsible for the 138 K low value of FM-$T_c$. Moreover, there are large proportions of $Ni^{3+}$-O-$Ni^{3+}$ and $Mn^{3+}$-O-$Mn^{3+}$ that lower the overall average local field and may explain in part this lower FM-$T_c$. In the ordered phase, a large value of FM-$T_c$ arises owing to the strong superexchange interaction in the $Ni^{2+}$-O-$Mn^{4+}$ bonds. The observed values of FM-$T_c$ for our films are also well consistent with the literature on bulk and thin films of LNMO samples. Past studies on LNMO, using various techniques [20] have clearly shown that the ordered phase includes only $Ni^{2+}$ and $Mn^{4+}$ oxidation states, while the disordered phase **is** due to the $Ni^{3+}$ and $Mn^{3+}$ ions. Our study further corroborates previous reports that the low temperature FM-$T_c$ is definitely related to the disordered phases of LNMO [10].

In the magnetic insulators, the spin-phonon coupling arises from the phonon modulation of the superexchange integral which **is** ultimately governed by the amplitude of the spin-spin correlation <$S_i.S_j$> functions where $S_i$ and $S_j$ are the localized spins at the $i^{th}$ and $j^{th}$ sites, respectively [17]. Within the mean-field framework, the phonon renormalization function $\delta\omega$ (T) [17] is related to the magnetization by $\delta\omega \propto M^2(T)/M_O^2$, where M(T) is the magnetization of the sample at a temperature T and $M_O$ is the magnetization at 0 K [17, 19]. To study the presence of spin-phonon coupling and their dependence on Ni/Mn ordering, we studied [19] the softening behavior of 672 cm$^{-1}$ stretching mode in the ordered films and their corresponding ones in the disordered and admixture phases. The net temperature dependence of the shifts [*i.e.,* $\delta\omega = \omega(300K) - \omega(T)$] in the stretching phonon mode are plotted in Fig. 3b. This figure reveals that irrespective of the Ni/Mn ordering, each film displays a phonon softening. However, their net magnitude differs significantly. In the disordered films it is only about 2.5 cm$^{-1}$, while in the ordered films it is about 6 cm$^{-1}$. Nonetheless, the softening in the ordered and disordered films begins at distinct temperatures and its onset point is very close to the magnetic transition temperatures. This reveals that a random distribution of Ni/Mn cation in LNMO disordered films partially prevents softening of this mode and suppresses the spin-phonon coupling. It also illustrates the presence of strong spin-phonon coupling in the ordered LNMO and a weak spin-phonon coupling in the disordered LNMO. In the case of the admixture films, we have been unable to measure the Raman spectra above 300 K due to the limitations of our set-up. Thus, it was not possible to identify unambiguously the softening onset point for the admixture films.

In summary, we have illustrated the influence of Ni/Mn ordering on the magnetic, phonon and spin-phonon coupling behavior in LNMO films. We also have elucidated the importance of growth conditions and their impact on Ni/Mn ordering in LNMO. Our study may open the possibility to promote and to control the cationic ordering and ultimately control the coupled behaviours of double perovskites for potential applications.

**Acknowledgments**

We thank S. Pelletier and M. Castonguay for their technical assistance. This work was supported by the Canadian Institute for Advanced Research, the Canada Foundation for Innovation, Natural Sciences and Engineering Research Council (Canada), Fonds Québécois pour la Recherche sur la Nature et les Technologies (Québec), and the Université de Sherbrooke.

**Figure captions**

**Figure 1: (Color online)** The phase-stability diagram of LNMO films grown by pulsed laser deposition. The circular area represents the suitable growth conditions to obtain Ni/Mn ordering, and the rectangular area represents the growth conditions suitable for obtaining disordered Ni/Mn configuration. Films with short-range ordered Ni/Mn are obtained in the rest of this diagram. The symbols position the growth conditions of our LNMO and some found in the literature: solid circles are for the present work; solid squares from Ref. 10 and yellow stars from Refs. 16. Top-left schematic illustrates the random distribution of Ni/Mn cations in the pseudo-cubic LNMO leading to a disordered unit cell, while the top-right shows the alternatively arranged Ni/Mn cations in an ideally ordered LNMO pseudo-cubic unit cell. [Red spheres (dark): B' cations; yellow spheres (gray): B'' cations, large spheres: A-cations; and blue spheres: oxygen atoms.] It is here important to note that the real crystal structure of the disordered phase is orthorhombic while the ordered phase is monoclinic. Arrows point out the narrow growth conditions used to stabilize these ordered and disordered phases.

**Figure 2: (Color online)** Comparative polarized Raman spectra of ordered, admixture and disordered films measured at 10 K in **(a)** the XX and **(b)** XY configurations.

**Figure 3: (Color online)** Comparative temperature dependence of **(a)** the magnetization and **(b)** the net softening of the 672 cm$^{-1}$ phonon in the disordered, admixture, and ordered films. The M-T curves were measured in a magnetic field of 500 Oe, and the softening in the ordered phase begins close to its magnetic transition temperature illustrating the presence of spin-phonon coupling in LNMO.

Fig. 1

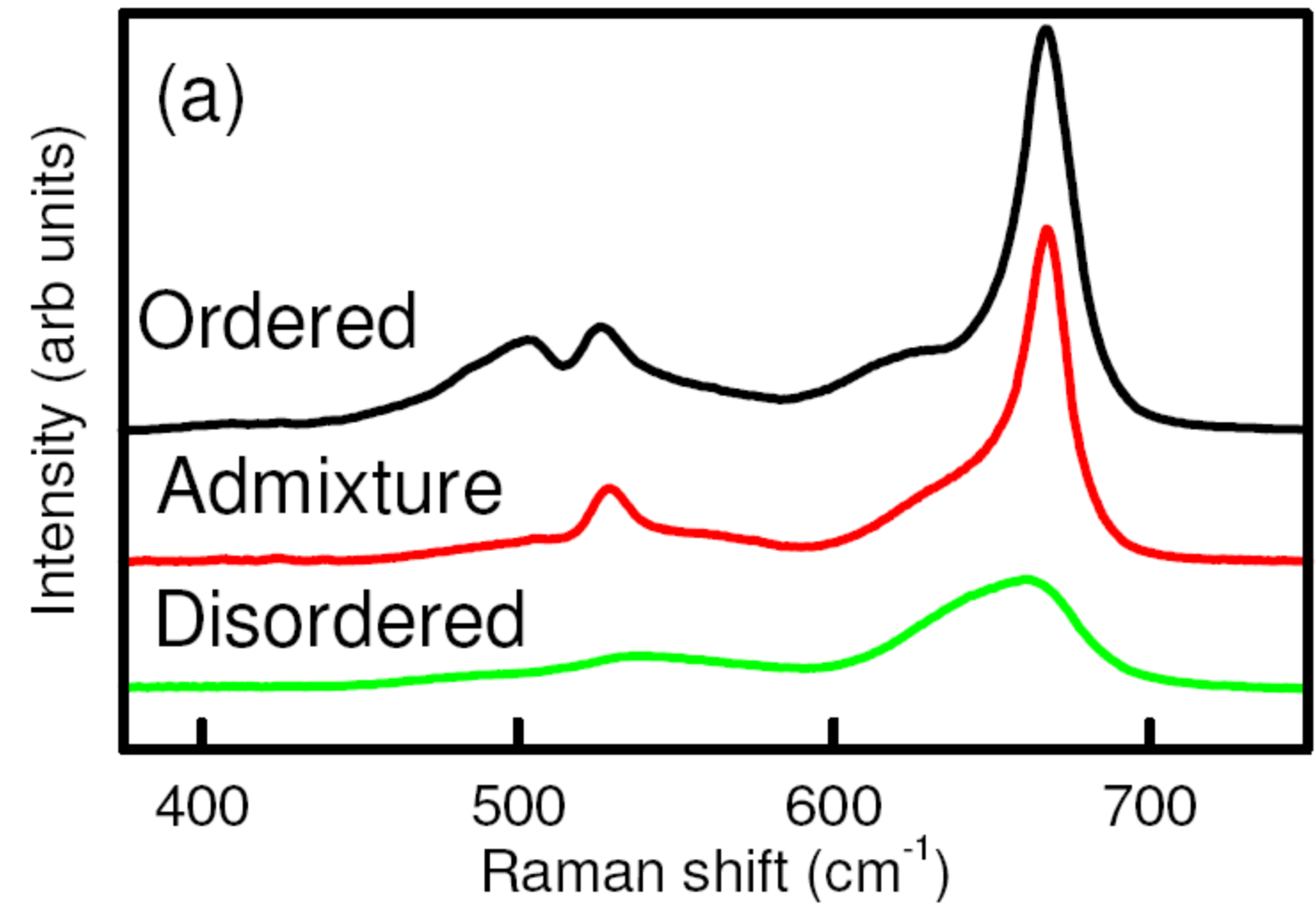

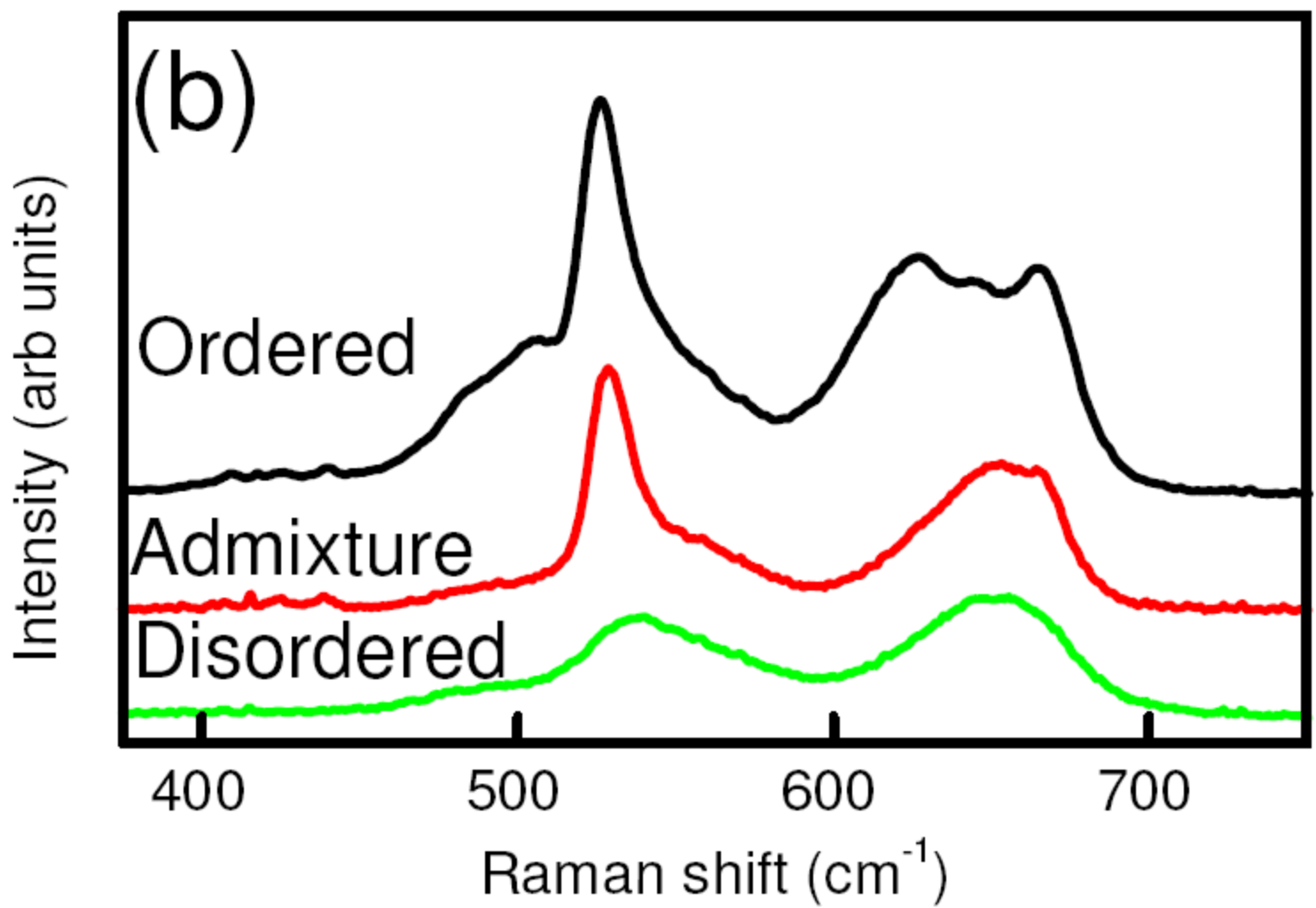

Fig. 2

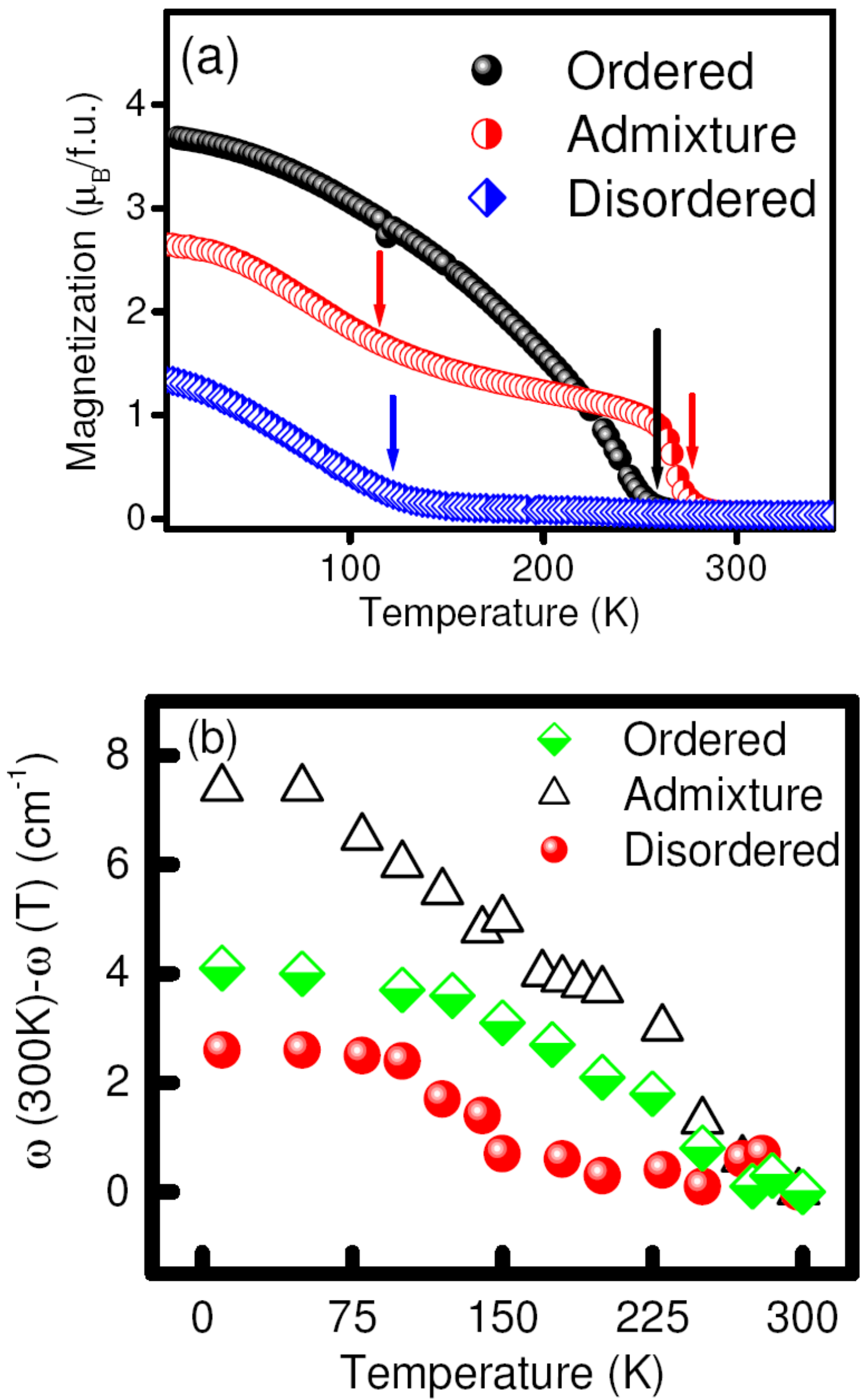

Fig 3